\documentclass[usegraphicx,usenatbib]{mn2e}
\title[Precession due to a close binary system]{Precession due to a close binary system: An alternative explanation for $\nu$-Octantis?}

\author[M.H.M. Morais and A.C.M. Correia]{M.H.M. Morais$^{1}$\thanks{E-mail: helena.morais@ua.pt} and A.C.M. Correia$^{1,2}$ \\
$^{1}$Department of Physics, I3N, University of Aveiro,
	     Campus Universit\'ario de Santiago, 3810-193 Aveiro, Portugal \\
$^{2}$ Astronomie et Syst\`emes Dynamiques, IMCCE-CNRS UMR 8028, 
             77 Avenue Denfert-Rochereau, 75014 Paris, France}

\begin{document}

\date{}

\maketitle

\label{firstpage}

\begin{abstract}

We model the secular evolution of a star's orbit when it has a nearby binary system. We assume a hierarchical triple system where the inter-binary distance is small in comparison with the distance to the star. We show that the major secular effect is precession of the star's orbit around the binary system's centre of mass. We explain how we can obtain this precession rate from the star's radial velocity data, and thus infer the binary system's parameters.  We show that  the secular effect of a nearby binary system on the star's radial velocity can sometimes mimic a planet.
We analyze the radial velocity data for $\nu$-octantis A which has a nearby companion ($\nu$-octantis B) and we obtain retrograde precession of $(-0.86\pm0.02)^\circ$/yr. We show that if $\nu$-octantis B was itself a double star, it could mimic a signal with similarities to that previously identified as a planet of $\nu$-octantis A.
Nevertheless, we need more observations in order to decide in favor of the double star hypothesis.

\end{abstract}

\section{Introduction}

Most extra-solar planet detections rely on measuring the parent star's wobble which is assumed to be caused by a planet. However, other effects can cause stellar wobble thus it is important to study these in order to avoid erroneous new planet announcements, as it already happened in the past \citep{Queloz2001A&A,Santos_etal2002A&A}.
 
In previous articles \citep{Morais&Correia2008,Morais&Correia2011}, we studied the short-term effect of a binary system on a star's motion. We saw that this could mimic a planet companion to the star  under some circumstances. In these articles, we considered moderately close binary systems ($\ge 10$~AU) in which the star's motion around the binary's centre of mass had periods of several decades.  Moreover, we realistically  assumed that we had observational data for only a fraction of this period, and not several orbits.

Here, we will present another scenario that requires a different analysis. We consider a star with a close binary system ($<5$~AU) and we assume that the observational data covers a few periods of the star's motion around the binary system's centre of mass. We will show that, in this case, we have to take into account secular effects which lead to slow precession of the star's orbit. 

In Sect.~2 we present the secular theory for hierarchical triple star systems composed of a star and a nearby binary. In Sect.~3 we show how we can measure the secular precession of the star's orbit from radial velocity data and how we can predict the binary system's parameters from this measurement. In Sect.~4 we apply the results from previous Sections to fictitious hierarchical triple star systems. In Sect.~5 we discuss the reported finding of a  planet  in the binary system $\nu$-Octantis. In Sect.~6 we present our conclusions.

\section{Secular theory for hierarchical triple star systems}

We consider a triple star system composed of an observed star, $m_2$, and a nearby binary of masses $m_0$ and $m_1$.
We use the Jacobi coordinates $\vec{r_1}$ (distance of $m_1$ to $m_0$), and
$\vec{r_2} $ (distance of $m_2$ to the centre of mass of $m_0$ and $m_1$). Moreover, we assume that $|\vec{r}_{1}|  \ll  |\vec{r}_{2}|$ (hierarchical triple system).

\subsection{Secular Hamiltonian}

The Hamiltonian is  \citep{Lee&Peale2003ApJ,Farago&Laskar2010}
\begin{equation}
H = -\frac{G m_{0} m_{1}}{2\,a_1} -\frac{G (m_{0}+m_{1}) m_{2}}{2\,a_2}+ F \ ,
\end{equation}
where the 1st term describes the Keplerian motion of $m_1$ with respect to $m_0$ (inner binary),
the 2nd term describes the Keplerian motion of $m_2$ with respect to the centre of mass of $m_0$ and $m_1$ (outer binary),
and
\begin{eqnarray}
F &=& -G m_{0} m_{2} \left(\frac{1}{r_{02}}-\frac{1}{r_2}\right) 
       -G m_{1} m_{2} \left(\frac{1}{r_{12}}-\frac{1}{r_2}\right) \ ,
\end{eqnarray}
where $G$ is the gravitational constant, the distance of $m_2$ to $m_0$ is
\begin{equation}
\vec{r}_{02} = \vec{r}_{2} + \frac{m_1}{m_{0}+m_{1}} \vec{r}_{1} \ ,
\end{equation}
and the distance of $m_2$ to $m_1$ is
\begin{equation}
\vec{r}_{12} = \vec{r}_{2} - \frac{m_0}{m_{0}+m_{1}} \vec{r}_{1} \ .
\end{equation} 
Expanding $1/r_{02}$ and $1/r_{12}$ in powers of $\rho=r_{1}/r_{2}$, and retaining terms up to order $\rho^2$,  we obtain the quadrupole Hamiltonian
\begin{eqnarray}
\label{quadrupole}
F &=&-\frac{G m_2}{r_{2}} \frac{m_0 m_1}{m_0+m_1} \frac{\rho^2}{2} \left( 3\,(\hat{r}_{1} \cdot \hat{r}_{2})^2-1 \right)  \ ,
\end{eqnarray}
where $\hat{r}_1$ and $\hat{r}_2$ are the versors of $\vec{r}_1$ and $\vec{r}_2$, respectively.

The secular quadrupole Hamiltonian is obtained by averaging Eq.~(\ref{quadrupole}) with respect to the inner and outer binary's orbital periods \citep{Farago&Laskar2010}
\begin{eqnarray}
\label{hamiltonian}
\bar{F} &=& C \left[ 2 -12\,e_1^2 - 6\,(1-e_1^2) (\hat{k}_{1} \cdot \hat{k}_{2})^2 \right. \nonumber \\
&& \left. + 30\,e_1^2 (\hat{i}_{1} \cdot {\hat{k}_2})^2 \right] \ ,
\end{eqnarray}
where 
\begin{equation}
\label{c}
C= \frac{G}{16}  \frac{m_0 m_1}{m_0+m_1}\frac{m_2}{(1-e_2^2)^{3/2}} \frac{a_1^2}{a_2^3} \ ,
\end{equation}
$\hat{k}_2$ and $\hat{k}_1$ are, respectively, the versors of the angular momentum vectors of outer binary ($\vec{G}_2$) and inner binary ($\vec{G}_1$), and $\hat{i}_1$ is the unit vector in the inner binary's orbital plane that points towards the inner binary's pericentre.

In an arbitrary reference frame we have
\begin{eqnarray}
\label{ci}
\hat{k}_{1} \cdot \hat{k}_{2} &=& \sin{I_1}\,\sin{I_2}\,\cos(\Omega_1-\Omega_2) +\cos{I_1}\,\cos{I_2} \\
\label{sisom}
\hat{i}_{1} \cdot \hat{k}_{2} &=&-\sin{I_2}\,\cos{I_1}\,\sin{\omega_1}\,\cos(\Omega_1-\Omega_2)  \nonumber \\
&& -\sin{I_2}\,\cos{\omega_1}\,\sin(\Omega_1-\Omega_2) \nonumber \\
&& +\sin{I_1}\,\sin{\omega_1}\,\cos{I_2} \  ,
\end{eqnarray}
with $I_1$ (inner binary's inclination), $I_2$ (outer binary's inclination), $\Omega_1$ (inner binary's  longitude of ascending node), $\Omega_2$ (outer binary's  longitude of ascending node), $\omega_1$ (inner binary's argument of pericentre).

\citet{Kozai1962AJ,Krymo&Mazeh1999MNRAS,Ford_etal2000ApJ} write the Hamiltonian in the invariant plane reference frame (Fig.~1). In this reference frame, $\Omega_1-\Omega_2=180^{\circ}$, and the relative inclination is $i=I_1+I_2$, thus
\begin{equation}
\label{hamiltonian0}
\bar{F}=C [(2+3\,e_{1}^2)(3\,\cos^2{i}-1)+15\,e_{1}^2\,\sin^2{i}\,\cos(2\,\omega_1)] 
\end{equation}
and due to conservation of angular momentum
\begin{equation}
\label{gsquared}
G^2=G_{1}^2+G_{2}^2+2\,G_1\,G_2\,\cos{i}=const \ .
\end{equation}

\begin{figure}
  \centering
    \includegraphics[width=8cm]{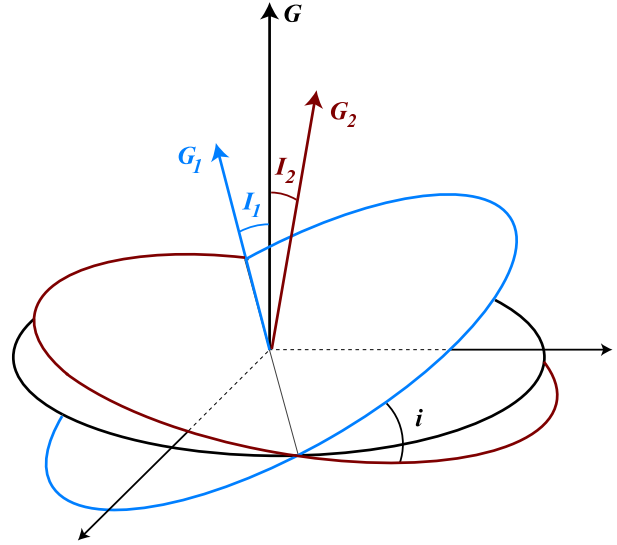}
  \caption{The invariant plane is orthogonal to the total angular momentum vector $\vec{G}=\vec{G}_1+\vec{G}_2$, where $\vec{G}_2$ and $\vec{G}_1$ are, respectively, the angular momentum vectors of the outer and inner binary's.}
  \end{figure}

\subsection{Secular equations}

The Delaunay canonical variables for this triple system are the  angles $l_j$ (mean anomalies), $\omega_j$ (arguments of pericentre), $\Omega_j$ (longitudes of ascending nodes), and their conjugate momenta, respectively
\begin{eqnarray}
L_j &=& \beta_j \sqrt{\mu_j a_j} \\
G_j &=& L_j \sqrt{1-e_j^2} \\
H_j &=& G_j \cos{I_j} \ .
\end{eqnarray}
with $j=1$ (inner binary) and $j=2$ (outer binary),
\begin{eqnarray}
\label{beta1}
\beta_1 &=&  \frac{m_0 m_1}{m_0+m_1}  \ , \\
\beta_2 &=& \frac{m_2 (m_0 + m_1)}{m_0+m_1+m_2} \ ,
\end{eqnarray}
$\mu_1=G (m_0+m_1)$ and $\mu_2=G (m_0+m_1+m_2)$.

By definition, the secular Hamiltonian does not depend on the mean anomalies, $l_j$, hence their conjugate momenta, $L_j$, and thus the semi-major axes, $a_j$, are constant.
The secular evolution is obtained from
\begin{eqnarray}
\label{gdot}
\dot{G}_j &=&  \frac{\partial{\bar{F}}}{\partial{\omega_j}} \ ,  \\
\label{omdot}
\dot{\omega}_j &=&  -\frac{\partial{\bar{F}}}{\partial{G_j}} \ , \\
\label{hdot}
\dot{H}_j &=& \frac{\partial{\bar{F}}}{\partial{\Omega_j}} \ , \\
\label{bigomdot}
\dot{\Omega}_j &=&  -\frac{\partial{\bar{F}}}{\partial{H_j}} \  .
 \end{eqnarray}
The quadrupole Hamiltonian (Eq.~\ref{hamiltonian}) does not depend on $\omega_2$ hence from Eq.~(\ref{gdot}) with $j=2$, $G_2$ and $e_2$ are constant.

\citet{Kozai1962AJ} and \citet{Kinoshita&Nakai2007} derived expressions for the secular evolution of the inner binary in the limit $m_1=0$ (inner restricted problem), using the quadrupole Hamiltonian. In this case, the outer binary's orbit  coincides exactly with the  invariant plane which is the natural choice of reference frame.
The  Hamiltonian is given by Eq.~(\ref{hamiltonian0})\footnote{When $m_1=0$ we must replace $\beta_1=1$ (Eq.~\ref{beta1}) into Eq.~\ref{c} \citep{Kozai1962AJ}. Note that there is a mistake in the expression given in \citet{Kinoshita&Nakai2007} since it should have $a_1^2$ and not $a_2^2$ in the nominator.} and the secular evolution of the inner binary i.e.\  $(e_1,\omega_1)$ are given by Eqs.~(\ref{gdot}) and (\ref{omdot}) with $j=1$.  Moreover, the secular oscillations of $e_1$ and $i$ are coupled due to conservation of angular momentum (Eq.~\ref{gsquared})\footnote{Since $G^2=G_1^2+G_2^2+2\,G_1\,G_2\,\cos{i}=const$, $G_2=const$ and $G_1\ll G_2$ hence $(1-e_1^2)\cos^2{i}\approx const$}.  

Following \citet{Kozai1962AJ} and \citet{Kinoshita&Nakai2007} we will describe the secular dynamics of prograde orbits ($0^\circ<i<90^\circ$) but since  Eq.~(\ref{hamiltonian0}) is invariant with respect to the transformation $i\rightarrow 180^\circ-i$, the secular dynamics is the same for prograde (inclination $i$) or retrograde (inclination $180^\circ-i$) orbits.
When $0<i<i_{c} \approx  40^{\circ}$, $\omega_1$ circulates while the eccentricity, $e_1$, and  relative inclination, $i$, exhibit secular oscillations with amplitude that increases with the relative inclination (in particular, coplanar orbits keep constant eccentricity, $e_1$). When $i>i_{c} \approx  40^{\circ}$, there are stationary solutions ($i=const$, $e_{1}=\sqrt{1-(5/3)\,\cos^2{i}}=const$, $\omega_{1}=\pm 90^{\circ}$), and  Kozai cycles  where  $i$, $e_1$ and  oscillate around the stationary solutions.

\citet{Farago&Laskar2010} showed that the secular motion of the inner binary when $0\le i\le 90^\circ$ (prograde orbits) is, in  the general problem ($m_1\ne 0$), equivalent to the inner restricted problem ($m_1=0$), as long as 
\begin{equation}
\label{limit0}
G_1^2-L_1^2+4\,G_2^2+2\,G_1\,G_2\,\cos{i} >0 \ .
\end{equation}
Defining $X=L_2/L_1$, we can write Equation~(\ref{limit0}) as
\begin{equation}
\label{limit}
4 \left(1-e_2^2 \right)\,X^2+2\,\sqrt{1-e_1^2}\,\sqrt{1-e_2^2}\,\cos{i}\,X-e_1^2 >0 \  .
\end{equation}
The left-hand side of Eq.~(\ref{limit}) is a second degree polynomial  in $X$ which is a convex function with two roots $X_1<0$ and $5/2\ge X_2>X_1$. Therefore, inequality (Eq.~\ref{limit0}) is verified if $L_1/L_2<2/5$ which is generally true if $a_1/a_2\ll 1$ (hierarchical system)  unless $m_0+m_1 \gg m_2$.

Equations~(\ref{gdot}) and (\ref{omdot}) with $j=2$ describe the secular evolution of the outer binary's eccentricity and argument of pericentre. 
We saw that the secular quadrupole Hamiltonian (Eq.~\ref{hamiltonian}) does not depend on 
$\omega_2$, hence from Eq.~(\ref{gdot}) with $j=2$, the conjugate momentum, $G_2$, and thus the eccentricity, $e_2$, are constant.  From Eq.~(\ref{omdot}) with $j=2$
we obtain the outer binary pericentre's precession rate
\begin{eqnarray}
\label{om2dotgen}
\dot{\omega}_2 & = & \frac{12\,C}{G_2}  \left[ \frac{1}{2} -3\,e_1^2 - \frac{3}{2}\,(1-e_1^2) (\hat{k}_{1} \cdot \hat{k}_{2})^2 \right. \nonumber \\
&& \left. + \frac{15}{2}\,e_1^2 (\hat{i}_{1} \cdot {\hat{k}_2})^2 \right] \ .
\end{eqnarray}

Higher order secular octupole terms cause long-term small amplitude oscillations in $e_1$ and $e_2$, which are more important for small to moderate values of the relative inclination, $i$ \citep{Krymo&Mazeh1999MNRAS,Ford_etal2000ApJ,Lee&Peale2003ApJ}.

\subsection{Precession of the outer binary's orbit}

The outer binary's precession rate (Eq.~\ref{om2dotgen}) depends on the secular motion of the inner binary. 
Moreover, 
in the invariant reference frame (Fig.~1) we have
\begin{equation}
\label{invariant}
 G_1\,\sin{I_1}=G_2\,\sin{I_2}
 \end{equation}
Since $a_1\ll a_2$ 
(hierarchical system) then, in general, $G_1\ll G_2$ (unless $m_0+m_1\gg m_2$) thus $\sin{I_2} \ll 1$ i.e.\ the outer binary's motion coincides approximately with the invariant plane.
Therefore, we express the right hand side of Eq.~(\ref{om2dotgen}) using the reference plane of the outer binary's orbit, i.e.\ setting $I_2=0$ and $I_1=i$ in Eqs.~(\ref{ci}),~(\ref{sisom}) thus obtaining
\begin{eqnarray}
\label{om2dotgen0}
\dot{\omega}_2 & \approx & \frac{12\,C}{G_2} A  \\
\label{atheta}
A &=& \left( \frac{1}{2}+\frac{3}{4}\,e_{1}^2 \right)(3\,\theta^2-1) \\ \nonumber
&& +\frac{15}{4}\,e_{1}^2(1-\theta^2)\cos(2\,\omega_1) \  ,
\end{eqnarray}
where $\theta=\cos{i}$.

Equation (\ref{om2dotgen0}) is an approximation of the precession rate, $\dot{\omega}_2$, because  the outer binary's orbit is not fixed but exhibits small amplitude oscillations around the invariant plane. 
The angle $\omega_1$ on the right hand side of Eq.~(\ref{atheta}) is measured with respect to the outer binary's orbit or, equivalently, with respect to the invariant plane\footnote{Since the intersection of inner and outer binary's orbits (line of nodes) is in the invariant plane (Fig.~1), then the angle $\omega_1$ is the same when measured with respect to the outer binary's orbit or with respect to the invariant plane.}. This formulation is necessary in order to describe the motion of the inner binary \citep{Kozai1962AJ,Kinoshita&Nakai2007,Farago&Laskar2010}. 
However, the angle $\omega_2$ represents the location of the outer binary's periapse with respect to the intersection with the observer's plane (when dealing with radial velocity data,  this is the plane orthogonal to the line of sight).

The long-term evolution of Kozai cyles (which exist if $i>i_{c}\approx 40^\circ$) was investigated by \citet{Eggleton&Kiseleva2001ApJ,Fabrycky&Tremaine2007,Wu&Murray2007}.
Typically, if $e_1$ becomes close to unity during a Kozai cycle, a combination of tidal evolution and relativistic effects will eventually disrupt the Kozai cycle and freeze the relative inclination, $i$. This will be followed by tidal damping of the semi-major axis, $a_1$, and eccentricity, $e_1$. 
The end state of a Kozai cycle that reaches $e_1\approx 1$ will be a tighter inner binary (smaller $\alpha=a_1/a_2$) on a circular orbit. Obviously, if $\alpha\ll 1$ then, as  $\dot{\omega}_2 \propto \alpha^2\,n_2$, the precession of the outer binary's orbit will be slow thus difficult to detect from observational data. On the other hand, orbits nearby the Kozai stationary solution are less prone to undergo tidal evolution and should keep the original value of $\alpha$.

We will, therefore, assume three scenarios for the inner binary's motion in the invariant plane reference frame :
\begin{itemize}
\item{$i<i_{c}$ where $\bar{e}_1$ and
$\bar{\theta}$ are average values of the secular oscillations in $e_1$ and $\theta$, respectively.  If $e_1\neq 0$ then $\omega_1$ circulates and we have, on average, $\bar{A}=(1/2+3\,\bar{e}_1^2/4)(3\,\bar{\theta}^2 -1)$.
If  $e_{1} \approx 0$ then $A \approx (3\,\theta^2 -1)/2$.} 
\item{$i>i_{c}$ but the inner binary's orbit was initially a high amplitude Kozai cycle that was circularized by tidal damping. In this case we also have $A=(3\,\theta^2 -1)/2$.}
\item{$i>i_{c}$ and the inner binary is at the Kozai stationary solution with $\omega_{1}=\pm 90^{\circ}$ and $\theta^2=3\,(1-e_1^2)/5$. In this case $A=-5\,(2\,\theta^2-1)(\theta^2-1)$. }
\end{itemize}
In all scenarios above the precession rate, $\dot{\omega}_2$ is approximately constant.

In Fig.~2 we plot the normalized precession rate, $A$ given by Eq.~(\ref{atheta}), when $e_1=0$ and at the Kozai stationary solution.  
We see that when $e_1=0$, precession is prograde when $i<54.73^\circ$ and retrograde when $i>54.73^\circ$. At the Kozai stationary solution, which exists only when $i>i_{c}\approx 40^\circ$, precession is retrograde when $i>45^\circ$.

Figure 2 shows the normalized precession rate, $A$, for $0^\circ<i<90^\circ$ (prograde orbits). However, we saw previously that Eq.~(\ref{hamiltonian0}) is invariant with respect to the transformation $i\rightarrow 180^\circ-i$, hence the precession rate (Eq.~ref{atheta}) is the same for prograde (inclination $i$) or retrograde (inclination $180^\circ-i$) orbits.

\begin{figure}
  \centering
    \includegraphics[width=8cm]{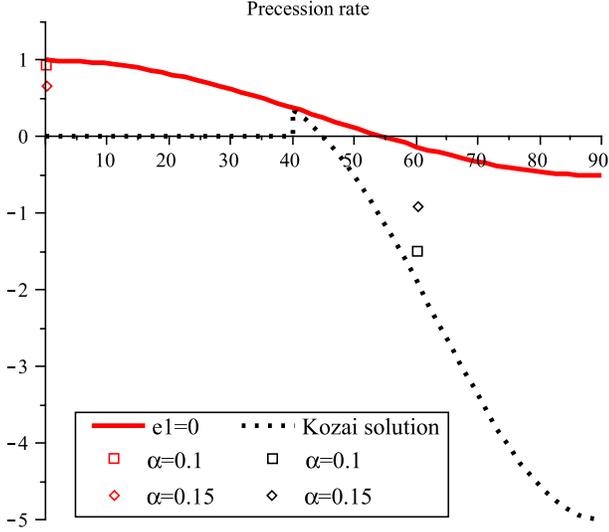}
  \caption{Comparison between theoretical precession rates ($A$ given by Eq.~\ref{atheta}) and values obtained in simulations.}
  \end{figure}

We performed numerical integrations of the equations of motion of  hierarchical triple star systems with parameters $m_2=M_{\odot}$, $a_2=3.0$~AU, $e_2=0.2$, and initial angles $I_2=0^{\circ}$, $\Omega_1=\Omega_2$ and 
$\omega_1=90^{\circ}$. 	The  inner binary had 
$m_1=0.08\,M_{\odot}$ and $m_0=0.42\,M_{\odot}$, semi-major axis ratio $\alpha=a_1/a_2$. We chose two configurations: (i) $i=0$ and  $e_1=0.01$ (coplanar nearly circular orbit);  (ii) $i=60^\circ$ and $e_1=0.76$  (Kozai stationary solution).
In Fig.~2 we show the comparison between the theoretical precession rates and the values obtained in the simulations
(up to 100~yrs). 
We see that the quadrupole approximation becomes  less accurate when we increase the semi-major axis ratio $\alpha$. This could be   due to truncation of the Hamiltonian at order $\alpha^2$, or it  could be due to inaccuracies of the first order secular theory \citep{Giuppone_etal2011AA}. 
These results do not vary much with other choices for the masses
$m_0$ and $m_1$, as long as $m_b=m_0+m_1$ is kept constant. 

\section{Orbital precession in radial velocity data}

\subsection{Measuring precession rates}

The radial velocity of a star, $m_2$, with a close binary system,  $m_b =m_0+m_1$, is approximately
\begin{equation}
V_r = V_{r0} + V_{rst}
\end{equation}
where $V_{rst}$ are short period perturbation terms obtained in \citet{Morais&Correia2011},
\begin{equation}
\label{radialvelocity}
V_{r0} = K ( \cos(f_2+\omega_2) + e_2 \cos(\omega_2 )) \ ,
\end{equation}
with
\begin{equation}
K = \frac{n_2 a_2}{\sqrt{1-e_{2}^2}} \frac{m_b}{m_{2}+m_b} \sin{I}_2 \ ,
\end{equation}
 and $\omega_2$ changes linearly with time (Sect.~2.3), i.e.
\begin{equation}
\label{omega2}
\omega_2 =\omega_{20} + \dot{\omega}_2 \,t \ .
\end{equation}

The precession of the star's orbit can be inferred from the radial velocity curve mostly due to the term
$\propto e_2$ in Eq.~(\ref{radialvelocity}). Typically, the observation timespan, $t_{obs}$, is much shorter than the precession cycle, thus if $\omega_{20}\neq 0,180^\circ$
\begin{equation}
K\,e_2\,\cos(\omega_2) \approx K\,e_2 [\cos(\omega_{20})-\sin(\omega_{20})\,\dot{\omega}_{2}\,t] 
\end{equation}
with $t\le t_{obs}$.

Hence,  Eq.~(\ref{radialvelocity}) is approximately a Keplerian radial velocity curve whose amplitude has a linear drift  which is at most ($\omega_{20}=90^\circ,270^\circ$)
 \begin{equation}
 \label{drift}
K\,e_2\,\dot{\omega}_{2}\,t_{obs} \ .
\end{equation}
If $\omega_{20}= 0,180^\circ$, the radial velocity curve's amplitude has  a quadratic drift $K\,e_2\,(\dot{\omega}_{2}\,t_{obs})^2/2$.

In order  to measure $\dot{\omega}_{2}$ with accuracy, two conditions must be met. First, the drift (Eq.~(\ref{drift})) must be larger than the observation's precision. Second, the observation timespan, $t_{obs}$, must be a few outer binary's periods so that
we can distinguish the secular drift, $\dot{\omega}_{2}\,t_{obs}$, from short period oscillations, $\Delta\omega$  \citep{Morais&Correia2011}, i.e.~we must have
$\dot{\omega}_{2}\,t_{obs}\gg \Delta\omega$ (see Fig.~3).
These two conditions help us predict when can we measure accurately the outer binary's precession rate. However, in practice we estimate $\dot{\omega}_{2}$ by fitting a precessing Keplerian orbit (Eqs.~\ref{radialvelocity} and \ref{omega2}) to the radial velocity data.

\begin{figure}
  \centering
    \includegraphics[width=8cm]{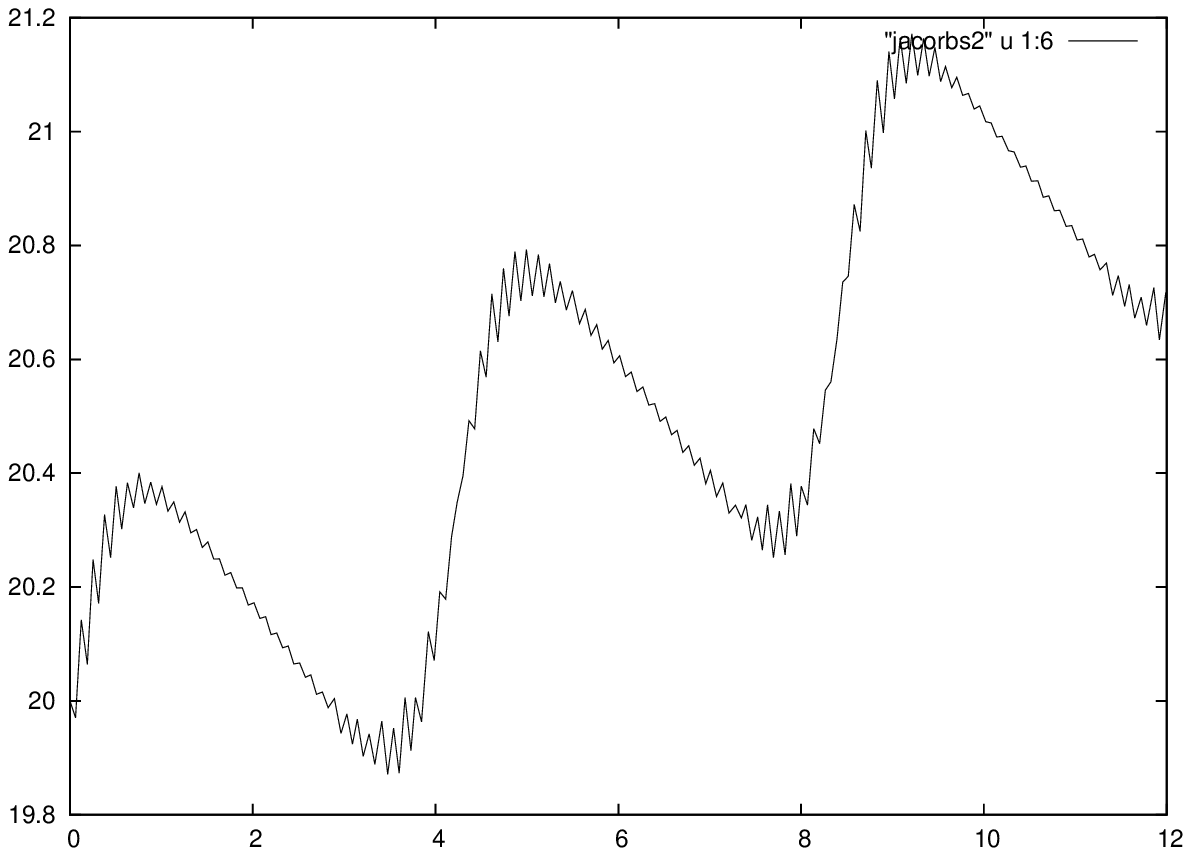}
    \includegraphics[width=8cm]{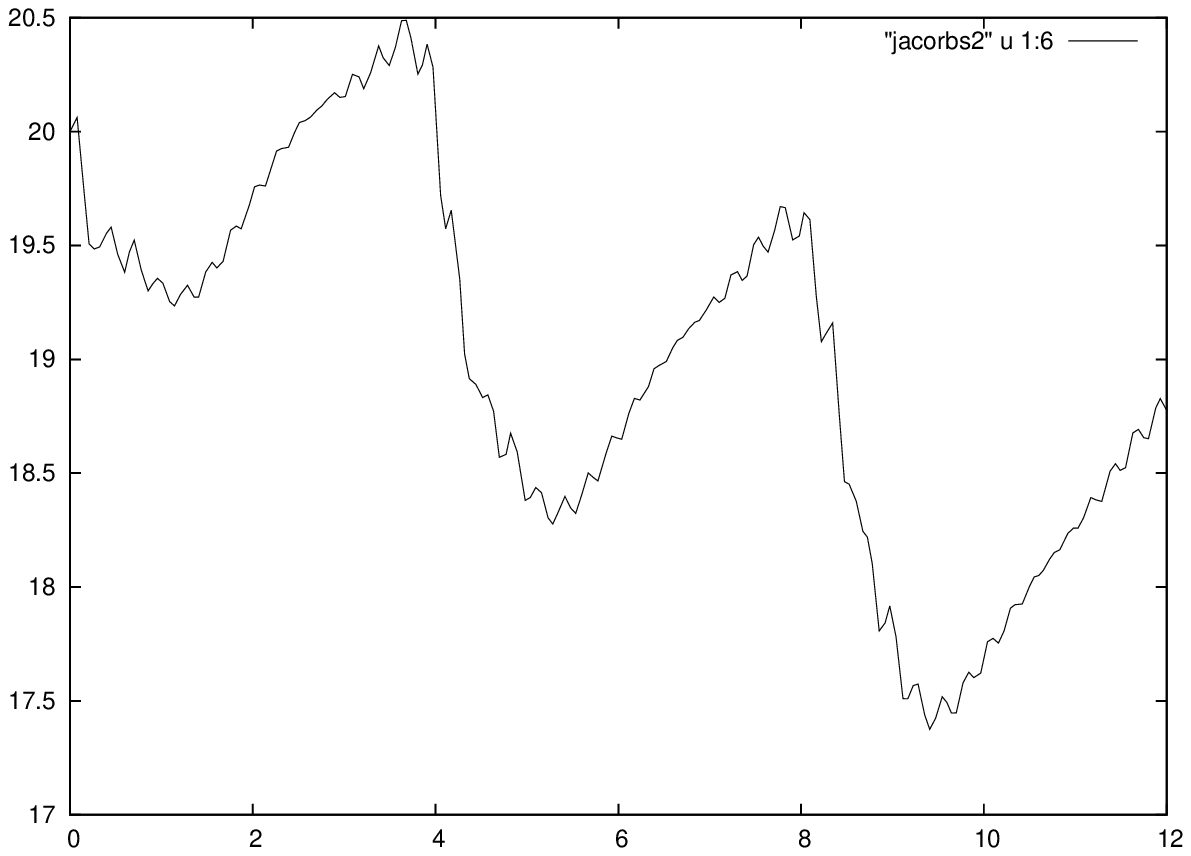}
    \put(-150,-10){time (years)}
\caption{Evolution of angle $\omega_2$, showing secular drift and short-period oscillations, obtained for simulations as in Sect.~4.1  but without observational errors: case I (top panel) and  case II (bottom panel).}    
\end{figure}

\subsection{Estimating inner binary parameters}

We saw that (Eqs.~(\ref{om2dotgen}) and (\ref{atheta}))
\begin{equation}
\label{precessionrate}
\dot{\omega}_{2} \propto  x (1-x) a_{1}^2 A \ .
\end{equation}
where $x=m_1/m_b$, $m_b=m_0+m_1$ and $A$ is a function of $\theta$ (cf.~Sect.~2.3).

Therefore, we can use the measured precession rate, $\dot{\omega}_{2}$, to estimate the parameters of an hidden binary 
(i.e.\  $m_1$ and $a_1$)  through the quantity $x (1-x)\,a_1^2$.
However, since $A$ is a function of  $\theta=\cos{i}$ which is unknown, we have to make some assumptions. From Fig.~2 in Sect.~2.3  we see that  prograde precession is faster when $i=0$ and that, if the inner binary is at the Kozai stationary solution, retrograde precession occurs  if $i>45^\circ$ and it is faster when $i=90^\circ$. As orbits with high relative inclination, $i$, will reach values of $e_1$ near unity, these are likely to become unstable or  undergo tidal evolution, hence we set a maximum value of $i=60^\circ$. Therefore, if $\dot{\omega}_2>0$ we assume that $i=0$ and $e_1=0$; 
if  $\dot{\omega}_2<0$
we assume $i=60^\circ$ and $e_1=0.76$. These assumptions imply maximum precession rates\footnote{However, if $e_1\neq 0$, the precession rate at $i=0$ increases by a factor $1+3\,e_1^2/2$ with respect to $e_1=0$.} (in absolute value), 
thus they give us minimum estimates for the parameter $x (1-x)\,a_1^2$. 

In order to estimate the hidden inner binary component's mass, $m_1$, we must provide an estimate for the inner binary's semi-major axis, $a_1$. 
\citet{Wiegert&Holman1999AJ} measured the size of stability regions around binary star
components. They assume a massless particle orbiting in the binary system's plane and find that this is stable if $a_1\le a_c$ with
 \begin{eqnarray}
 \label{acritical}
 \frac{a_c}{a_2} &=&  (0.464\pm0.006) + (-0.380\pm0.010)\,\mu   \nonumber \\
&& + (-0.631\pm0.034)\,e_2  +(0.586\pm0.061)\,\mu\,e_2 \nonumber \\ 
&&+(0.150\pm0.041)\,e_2^2 +(-0.198\pm0.074)\,\mu\,e_2^2 
 \end{eqnarray}
 where $a_2$ and $e_2$ are, respectively, outer binary's semi-major axis and eccentricity, and mass parameter $\mu=m_2/(m_2+m_b)$.

\section{Examples}

To test our model, we performed numerical integrations of the equation of motion of  hierarchical triple systems composed of a star, $m_2$, and a close by binary, $m_b$.
We chose masses $m_2=M_{\odot}$ and $m_b=0.5\,M_{\odot}$, and semi-major axes $a_2=3$~AU and $a_1=0.3$~AU,
which imply periods $T_2=4.24$~yr for the outer binary and $T_1=84.9$~day for the inner binary,  and semi-major axis ratio 
$\alpha=a_1/a_2=0.1$. 
 The initial angles were $I_2=90^{\circ}$, $\Omega_1=\Omega_2=0$, $\omega_1=90^{\circ}$, $\omega_2=20^\circ$.
We computed the radial velocity of the star $m_2$, and simulated observational data points for a timespan, $t_{obs}$, and a certain precision limit. We then applied the traditional techniques used in radial velocity data analysis.

\subsection{Measuring precession rates}

To test in which circumstances we are able to measure the outer binary's precession rate we set the outer binary on an eccentric orbit ($e_2=0.2$). 

Case I is a coplanar triple system ($i=0$ i.e. $I_1=I_2=90^{\circ}$) where the inner binary has masses $m_0=0.42\,M_{\odot}$ and $m_1=0.08\,M_{\odot}$,
and a nearly circular orbit ($e_1=0.01$).   
In Table~1 we show the results of fitting a  fixed Keplerian orbit (fit 0) or a precessing Keplerian orbit (fit 1) to the data time series with 128 points over $t_{obs} \approx 8$~yr, at precisions of about $5$~m/s (A) and $1$~m/s (B), respectively.
From Fig.~3 (top) we see that since $t_{obs}=8$~yr, $\Delta\omega>\dot{\omega}_2\, t_{obs}$.
The maximum radial velocity drift over $t_{obs}$ is $18.6$~m/s. At precision $5.425$~m/s (A) the observation error is $29\%$  of the maximum drift. Therefore, fit (1) is only slightly better than fit (0). However, at  precision $1.085$~m/s (B) the observation error is only $6\%$ of the maximum drift,  thus fit (1) is clearly better than fit (0). The theoretical value (quadrupole approximation) for the precession rate is $\dot{\omega}_2=0.093^\circ$/yr while the true value (simulations up to 100~yrs) is $\dot{\omega}_2=0.089^\circ$/yr.

\begin{table}
\centering
\begin{tabular}{|c||c|c|c||}
\hline
Case I & fit    & A & B \\
\hline \hline
prec (m/s) & -  & 5.425 & 1.085 \\ \hline \hline
$T$ (day) & (1) &$1544.12\pm0.05$ &   $1544.15\pm0.01$  \\
            & (0) &  $1544.15\pm0.05$ & $1544.18\pm0.01$  \\
\hline
$K$ (m/s) & (1) & $7172.9\pm0.7$ & $7173.86\pm0.13$  \\
                & (0) &  $7172.6\pm 0.7$ & $7173.62\pm0.13$  \\
\hline
$e$ & (1) & $0.19879\pm0.00009$ & $0.19873\pm0.00002$  \\
       & (0) & $0.19875\pm0.00009$ & $0.19870\pm0.00002$   \\
\hline
$\dot{\omega}$ ($^\circ$/yr) &  (1) & $0.115\pm0.016$ & $0.0937\pm0.0032$ \\        
\hline \hline
$\sqrt{\chi^2}$ & (1) & 1.569  &  1.660  \\
                        & (0) & 1.690  &  3.097  \\
\hline
$rms$ (m/s) & (1) & 9.1850  & 1.9154   \\
                    &  (0) & 9.8849 & 3.4024   \\
\hline \hline
\end{tabular}
\caption{Fits to Case I ($m_0=0.42\,M_{\odot}$, $m_1=0.08\,M_{\odot}$, $i=0$, $e_1=0.01$)
with $t_{obs}=8$~yr and different precisions.}
\end{table}

Case II has an inner binary with masses $m_0=0.35\,M_{\odot}$ and $m_1=0.15\,M_{\odot}$, with  $i=60^{\circ}$ (i.e.~$I_1=30^{\circ}$ and $I_2=90^\circ$) and $e_1=0.76$ (Kozai stationary solution).
In Table~2 we show the results of fitting a  fixed Keplerian orbit (fit 0) or a precessing Keplerian orbit (fit 1) to  data time series with 99 points over $t_{obs} \approx 6$~yr (C), and  to  data time series with 154 points over $t_{obs} \approx 12$~yr (D), both  at precision of about $5$~m/s.
The maximum radial velocity drifts over 6~yr and 12~yr are, respectively, $31$~m/s and $62$~m/s, which correspond to observation errors of, respectively, $16\%$  and $8\%$ of the maximum drift. From Fig.~3 (bottom) we see that when $t_{obs}=6$~yr, $\Delta\omega\approx \dot{\omega}_2\, t_{obs}$ while when  $t_{obs}=12$~yr, $\Delta\omega\ll \dot{\omega}_2\, t_{obs}$. Therefore, when  $t_{obs}=6$~yr (C),  fit (1) is slightly better than fit (0) but when $t_{obs}=12$~yr (D),  fit (1) is clearly better than fit (0). 
The theoretical value (quadrupole approximation) for the precession rate is $\dot{\omega}_2=-0.272^\circ$/yr  while the true value (simulations up to 100~yrs) is $\dot{\omega}_2=-0.238^\circ$/yr.

\begin{table}
\centering
\begin{tabular}{|c||c|c|c||}
\hline \hline
Case II & fit & C & D \\ \hline \hline
$t_{obs}$ (yr)  & - & $6$ & $12$  \\
\hline \hline
$T$ (day) &  (1) & $1567.74\pm0.07$ & $1567.75\pm   0.03$ \\
            &  (0) & $1567.44\pm0.07$  & $1567.70\pm0.03$  \\
\hline
$K$ (m/s) & (1) & $7135.81\pm0.89$ & $7135.61\pm0.60$   \\
                  &  (0) & $7139.36\pm0.84$  & $7134.50\pm0.60$  \\
\hline
$e$ &  (1) & $0.20495\pm0.00011$ &  $0.20504\pm0.00009$   \\
       &  (0) & $0.20491\pm0.00011$ & $0.20558\pm0.00009$   \\
\hline
$\dot{\omega}$ ($^\circ$/y) & (1) & $-0.256\pm0.021$ & $ -0.2606\pm0.0086$ \\
\hline \hline 
$\sqrt{\chi^2}$ & (1) & 1.656  &  1.628  \\
                        &  (0) & 2.077  &  2.984  \\
\hline
$rms$ (m/s) &  (1) & 9.4952 & 9.5135   \\
                    &  (0) & 11.6178 & 16.1299   \\
\hline \hline
\end{tabular}
\caption{Fits to Case II ($m_0=0.35\,M_{\odot}$, $m_1=0.15\,M_{\odot}$, $i=60^\circ$, $e_1=0.76$) with precision 5.4~m/s and different $t_{obs}$.}
\end{table}

\subsection{Estimating inner binary parameters}
 
We simulated triple systems as described above with the observed star on an eccentric orbit with $e_2=0.2$  (outer binary) around an inner binary with masses  $m_0=0.42\,M_{\odot}$ and $m_1=0.08\,M_{\odot}$, and semi-major axis $a_1=0.3$~AU. The values of the relative inclination were $i=0^\circ$,~$20^\circ$,~$40^\circ$,~$50^\circ$,~$55^\circ$,~$60^\circ$. When $i<40^\circ$ the inner binary had a nearly circular orbit ($e_1=0.01$) and when $i\ge 40^\circ$ it had an eccentric orbit near the Kozai stationary solution\footnote{The Kozai stationary solution has $e_{1}=\sqrt{1-(5/3)\,\cos^2{i}}$ and $\omega_1=\pm90^\circ$.}.
In all cases $t_{obs}=8$~yr, and the precision limit was about $1$~m/s.

In Table~3 we present the precession rates (theoretical and measured in the simulations) and the ratio between $\sqrt{\chi^2}$ of fit (1) and fit (0) which measures the goodness of fit (1) with respect to fit (0). As explained in Sect.~3.2, we obtain minimum estimates for the inner binary parameter, $x (1-x) a_1^2$ with $x=m_1/m_b$ (Eq.~\ref{precessionrate}), assuming $i=0$ and $e_1=0$ when $\dot{\omega}_2>0$, or $i=60^\circ$ and $e_1=0.76$ when  $\dot{\omega}_2<0$ (cf.~Fig.~2).  We can then obtain minimum estimates for $m_1$ assuming $a_1=a_c=0.49$~AU which is the maximum size of stable orbits (massless particle $m_1\ll m_b$) around $m_0\approx m_b$ in the coplanar case (see Eq.~(\ref{acritical}) with $e_2=0.2$ and $\mu=0.67$). 
These estimates are all realistic (minimum mass of the hidden inner binary companion between $6\,M_J$ and $27\,M_J$) hence  can be used as input parameters for a N-body fit which can provide best-choice values for $m_1$ and $a_1$.

\begin{table}
\centering
\begin{tabular}{|c||c||c|c|c|c|c|}
\hline
   i ($^\circ$) &  $e_1$ & $\dot{\omega}_2$ (t) &  $\dot{\omega}_2$ (s) &  ratio $\sqrt{\chi^2}$ & $m_1$ ($M_{\odot}$) \\
\hline
$0$ & 0.01 & 0.0928 & $+0.094$ &  0.54  & 0.027 \\
       &         &             & $\pm0.003$ &           &           \\
\hline
$20$ & 0.01 & 0.0765 & $+0.078$ & 0.60 & 0.022 \\
         &         &             & $\pm0.003$ &          &          \\
\hline
$40$ & 0.01 & 0.0353 &  $+0.040$ & 0.85 & 0.011 \\
         &         &             &  $\pm0.003$ &         &         \\
\hline
$50$ & 0.56 & -0.0473 &  $-0.040$  & 0.89  & 0.006 \\
         &          &             &  $\pm0.003$    &          &          \\     
\hline
$55$ &  0.67 & -0.1065 & $-0.099$ &  0.67  & 0.015 \\
         &          &              &  $\pm0.003$ &            &          \\
\hline
$60$ &  0.76 & -0.1740 & $-0.167$  & 0.50  & 0.025 \\
         &           &              & $\pm0.003$   &          &          \\
\hline
\end{tabular}
\caption{Hierarchical triples system ($\alpha=0.1$) with inner binary on nearly circular orbit (if $i<40^\circ$) or at Kozai stationary solution (if $i>40^\circ$): comparison between outer binary's theoretical (t) and observed (s) precession rates; $\sqrt{\chi^2}$ ratio between fit (1) and fit (0) to outer binary's orbit; minimum mass of hidden inner binary component, $m_1$.}
\end{table}

\subsection{Can precession mimic a planet?}

Here, we repeat the question already made in \citet{Morais&Correia2008,Morais&Correia2011}. If we do not know about the inner binary's presence because one of its components is unresolved, can the binary's effect be mistaken as a planet?

In \citet{Morais&Correia2008,Morais&Correia2011} we showed that the residuals leftover from fitting a fixed Keplerian orbit to the outer binary (observed star's orbit around the inner binary's centre of mass) contained additional periodic signals that could be mistaken by planets. Here, we show an example of similar behavior obtained from the previous simulation (case II with $t_{obs}=6$~yr) in Fig.~4 (top). The periodogram\footnote{We compute Generalized Lomb-Scargle periodograms as defined in \citet{Zechmeister&Kurster2009}} has an obvious peak at 606~day which is nearly commensurate (ratio 2/5) with the outer binary's period. However, this peak disappeared when we fitted a precessing Keplerian orbit to the outer binary (Fig.~4: middle), and it was no longer prominent when we increased the observation timespan to $t_{obs}=12$~yr (Fig.~4: bottom). 

\begin{figure}
  \centering
    \includegraphics[width=8cm]{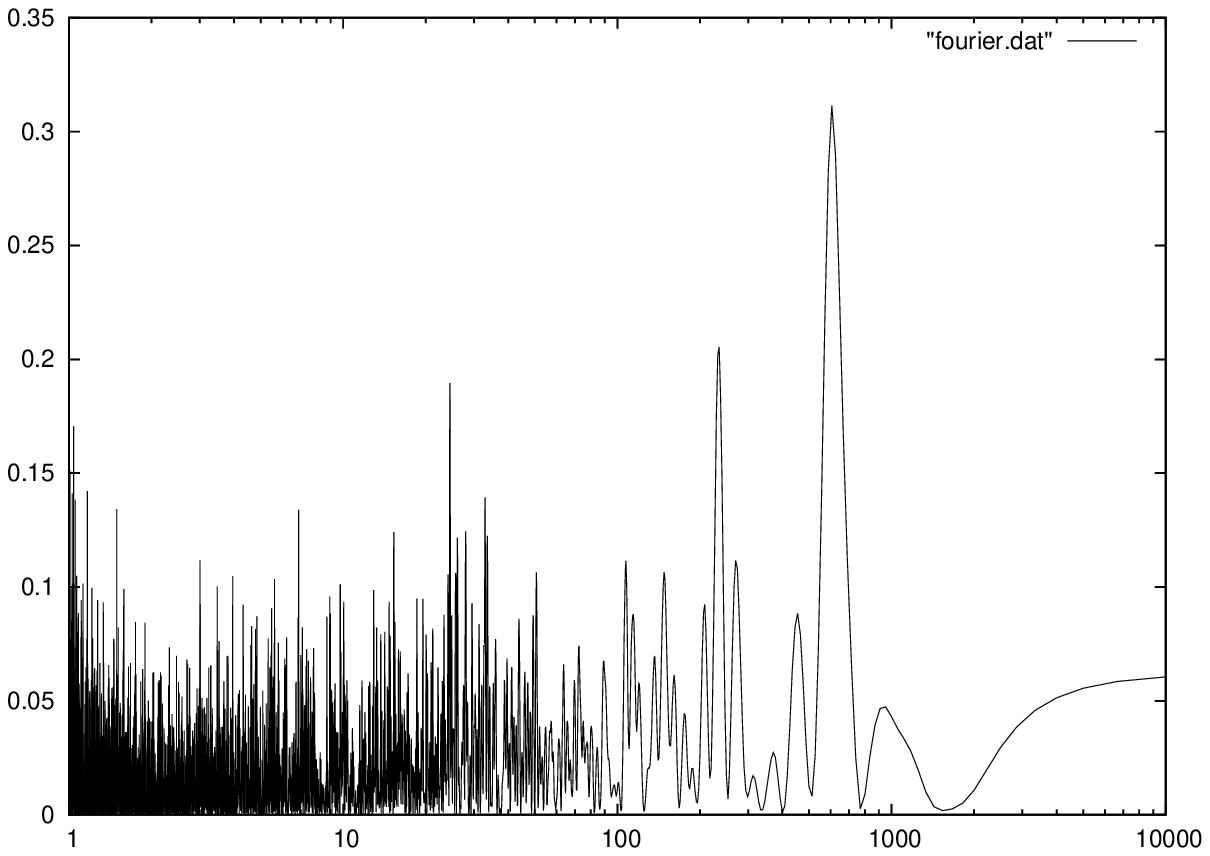}
    \includegraphics[width=8cm]{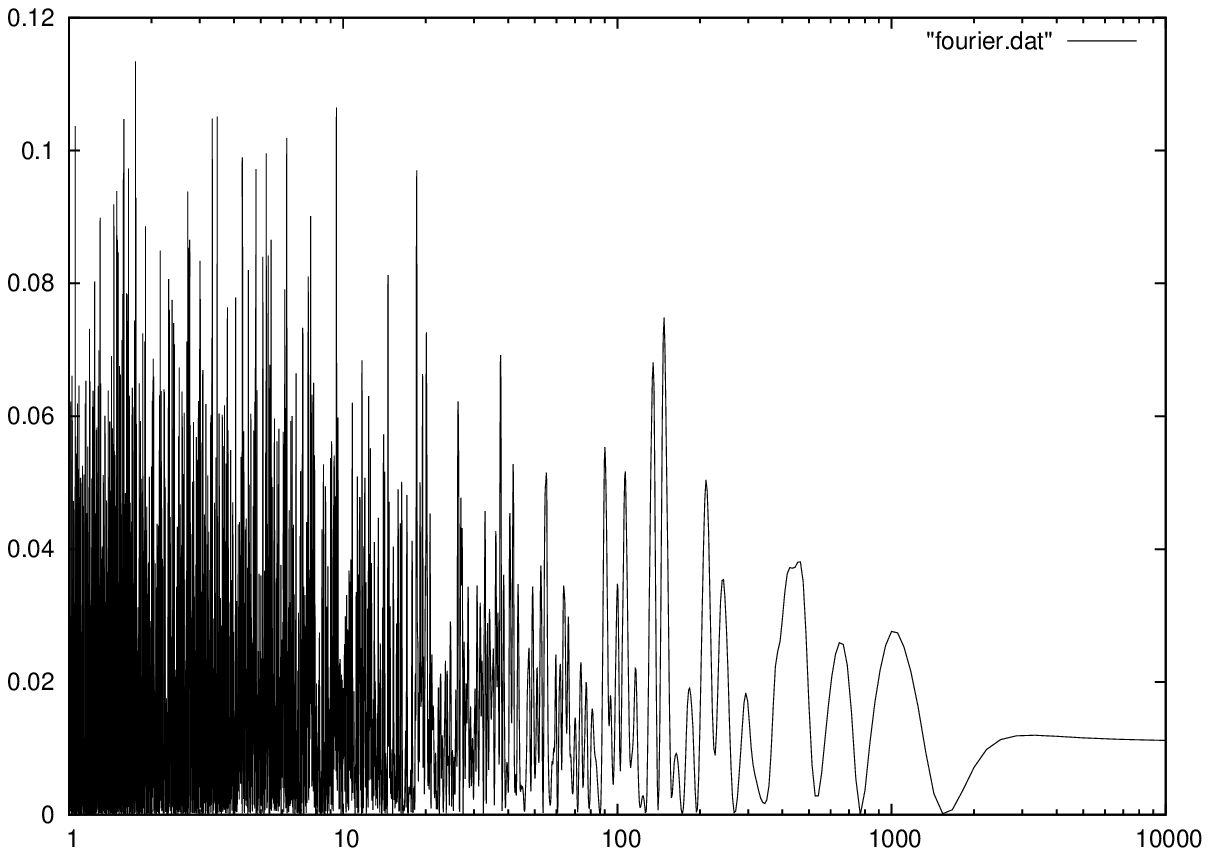}
    \includegraphics[width=8cm]{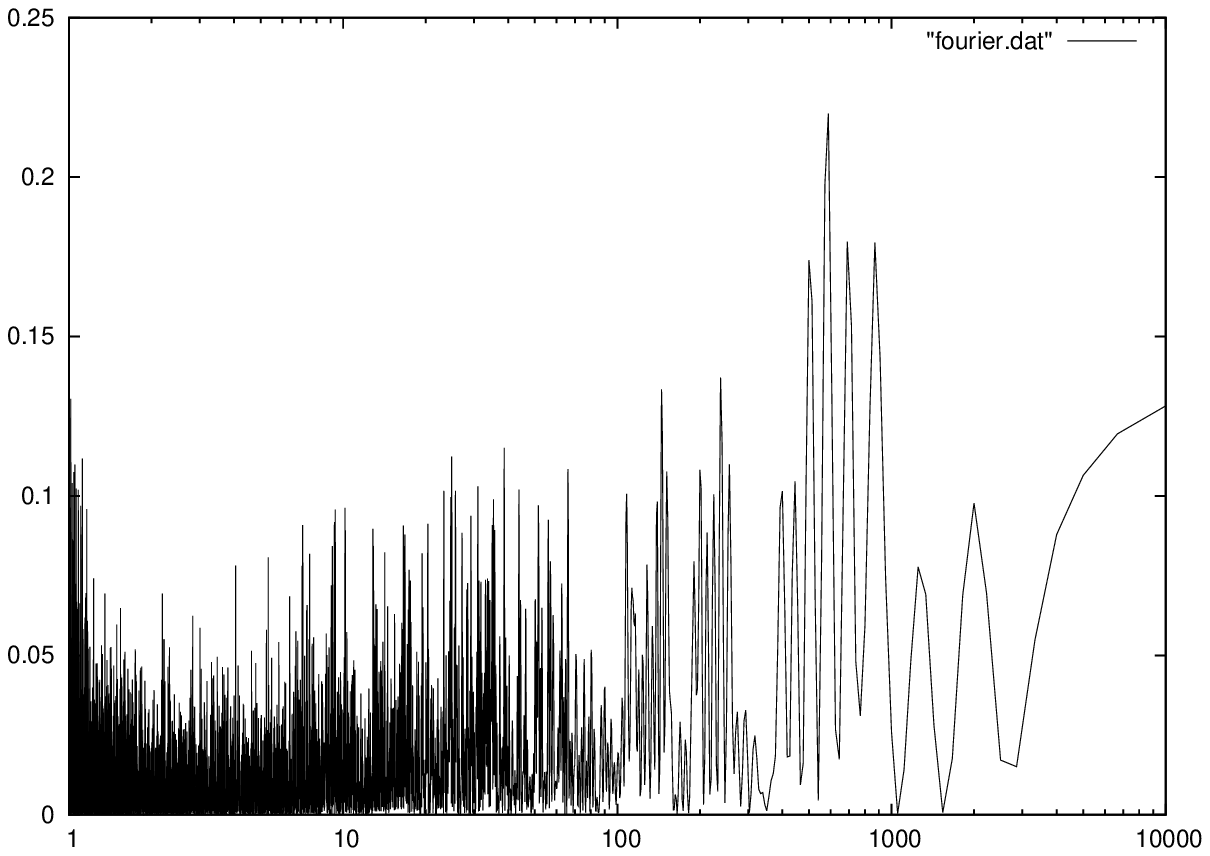}    
  \caption{ Periodogram of residuals after fit (0) and fit (1) to radial velocity data from simulation in Table~2 with $t_{obs}=6$~yr (top and middle, respectively) and after fit (0) to simulation in Table~2 with $t_{obs}=12$~yr (bottom).}
  \end{figure}
  
In \citet{Morais&Correia2011} we studied the short term effect of a binary system on a nearby star in the case of eccentric and inclined orbits. We saw that, when the observed star's orbit (outer binary) was eccentric, the radial velocity was composed of a main Keplerian term that described the star's motion around the inner binary's centre of mass, and short-period terms (obtained by integrating with respect to time Eq.~(34) in \citet{Morais&Correia2011}). In particular, some short-period terms appeared at harmonics of the outer binary's frequency, $n_2$.
 However, in \citet{Morais&Correia2011} we considered that $t_{obs}$ was only a fraction of the outer binary's period. In this situation, the outer binary's orbit was not well constrained and these harmonics were incorporated into the main Keplerian term. Now, as $t_{obs}$ covers a few outer binary's orbits, these harmonics will appear in the residuals leftover from fitting a fixed Keplerian curve to the outer binary. However, the observed star's orbit (outer binary) is, in fact, precessing and the  precession rate, $\dot{\omega_2}$, is approximately constant in the long-term but has short-term oscillations (see Fig.~3). If $t_{obs}$ covers only a few outer binary's orbits, these short term oscillations will cause mixing up of the frequencies thus the signals do not exactly coincide with harmonics of $n_2$. In particular, combinations of these harmonics can appear (Fig.~4: top). 
When we fit a precessing Keplerian orbit to the outer binary, as $t_{obs}$ is short compared to the precessional period, the signals at or nearby harmonics of $n_2$  are incorporated into the precessing Keplerian orbit (Fig.~4: middle).
As $t_{obs}$ increases, the short period oscillations become negligible with respect to the secular terms (Fig.~4: bottom).

If the observed star  has a circular orbit then there is no pericentre precession. When the inner binary's orbit is also circular but inclined ($i<40^\circ$) with respect to the outer binary\footnote{Due to the Kozai effect,  when $i>40^\circ$ the inner binary's orbit cannot remain circular.},  the star's radial velocity is composed of a main Keplerian term (circular orbit with frequency $n_2$) and short period terms (obtained by integrating with respect to time Eq.~(21) in \citet{Morais&Correia2011}). In particular, there are signals at frequencies $n_2$ and $3\,n_2$. The term with frequency $n_2$ is simply incorporated into the main Keplerian fit. However, the term with frequency $3\,n_2$ can be mistaken as a planet at the 3/1 mean motion resonance with a companion "star" of mass $m_b=m_0+m_1$.  

We simulated a triple system, as explained at the beginning of Sect.~4, with $m_1=m_2=0.25\,M_{\odot}$, $e_2=0$, $e_1=0$ and $i=30^\circ$. We generated radial velocity data with 154 points over $t_{obs}=12$~yr, at precision 0.543~m/s.  In Fig.~5 we show a periodogram of the residuals leftover after fitting a Keplerian orbit to the outer binary. As expected, we see peaks at 505~day (frequency $3\,n_2$, harmonic of $n_2$) and 46~day (frequency $2\,n_1-3\,n_2$, short-period term as in \citet{Morais&Correia2008})  with amplitudes 1.7~m/s and 1.2~m/s, respectively. These can be mistaken as planets.

\begin{figure}
  \centering
    \includegraphics[width=8cm]{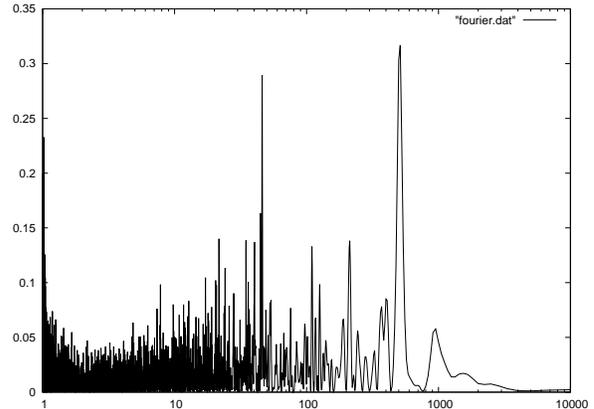}
\caption{Periodogram of residuals after fit (0) to  radial velocity data from simulation of circular non-coplanar triple star system composed of an observed star and a companion unresolved binary.
Peaks at 46 day and 505 day can be mistaken by planets.}    
\end{figure}

\section{A planet in $\nu$-Octantis?}

The system $\nu$-Octantis is a close single-line spectroscopic binary. \citet{Ramm_etal2009}  published  radial velocity data consisting of 221 points covering a timespan $t_{obs}=1862$~day, with precision around 5~m/s  (inferred from the published observation errors). Combining the radial velocity data with astrometric measurements,   \citet{Ramm_etal2009} 
derived improved parameters for  the $\nu$-Octantis binary (Table 4). The residuals leftover from fitting a Keplerian orbit to $\nu$-Octantis  show  an additional signal, which  \citet{Ramm_etal2009} identify as a planet  with minimum mass, $2.5\,M_J$, and semi-major axis, $1.2$~AU, orbiting the primary star (cf.~Table 4). Moreover, the planet's period, $417$~day, is nearly commensurate (ratio 2/5) with $\nu$-Octantis binary system's period.
 
\begin{table}
\centering
\begin{tabular}{|c|c|}
\hline \hline
\multicolumn{2}{c}{$\nu$-Octantis A}  \\ \hline 
\multicolumn{2}{c}{$m_A=(1.4\pm0.3)\,M_{\odot}$} \\ \hline \hline
\multicolumn{2}{c}{$\nu$-Octantis B + planet}  \\ \hline       
$T_2=(1050.11\pm0.13)$~day & $T_p=(417\pm4)$~day \\
$K_2=(7032.3\pm2.6)$~m/s & $K_p=(51.8\pm1.6)$~m/s \\
$e_2=0.2359\pm0.0003$ & $e_p=0.123\pm0.037$ \\
$\omega_2=(75.05\pm0.075)^\circ$ & $\omega_1=(260\pm21)^\circ$ \\
$I_2=(70.8\pm0.9)^\circ$ & $I_p=?$ \\
$\Omega_2=(87\pm1)^\circ$ & $\Omega_p=?$ \\ 
\hline
$m_B=(0.5\pm0.1)\,M_{\odot}$ &   $m_p\,\sin{I_p}=2.5\,M_J$ \\   
$a_2=(2.55\pm0.13)$~AU & $a_p=(1.2\pm0.1)$~AU \\    
\hline 
\multicolumn{2}{c}{$\sqrt{\chi^2}=4.2$} \\
\multicolumn{2}{c}{$rms=19$~m/s} \\
\hline \hline
\end{tabular}
\caption{Fitted parameters for $\nu$-Octantis and possible planet 
 \citep{Ramm_etal2009}.}
\end{table}
 
A planet about half way between the primary and secondary stars (semi-major axis ratio  $\alpha=0.47$) is unexpected. In fact, according to \citet{Wiegert&Holman1999AJ} the stability limit for coplanar prograde orbits around the primary star  is only $a_c=0.6$~AU (Eq.~\ref{acritical}). Nevertheless, \citet{Eberle&Cuntz2010} propose that such planet ($a_p=1.2$~AU) can be stable for at least 10 million years on a retrograde coplanar orbit. 

In Table~5 we present the results of fitting the radial velocity data  from \citet{Ramm_etal2009}  with a fixed Keplerian orbit (fit 0) and 
a precessing  Keplerian orbit (fit 1). We see that fit (1) with $\dot{\omega}_2=-0.86\pm0.02^\circ/yr$ is better than fit (0),
although the difference is not yet very significant. However, as seen previously in Sect.~4.1, this could be due to the short observation timespan ($t_{obs}=1.77\,T_2$). In Fig.~6 we show the periodogram of the residuals after fit (0) and fit (1). After fit (0) there is a prominent signal at 417~day, while after fit (1) this signal is still present but it is no longer dominant and seems to be within the noise level.  This is similar to the behavior described in Sect.~4.3. From Table~5 we  see that although fit (1) is better than fit (0), the fit with a planet at 417~day is currently better. As explained above, this could be due to the short-observation timespan, or even due to the particular  sampling of the radial velocity data.
 Moreover, the fit with the planet introduces additional five free parameters while  fit (1) introduces only one more free parameter ($\dot{\omega}_2$) which  could also help explain why the fit with the planet seems better than fit (1).

\begin{table*}
\centering
 \begin{minipage}{120mm}
\begin{tabular}{|c||c|c|c|c||}
\hline \hline
 & fit  & $\nu$-Octantis & simulation I & simulation II  \\
\hline \hline
 & (1) & $1050.46\pm0.03$  &  $1114.63\pm0.04$  & $1093.44\pm0.03$ \\  
$T$ (day)       & (0) & $1050.11\pm0.03$   &  $1114.32\pm0.03$ & $1093.04\pm0.03$   \\ 
            & (0)+pl  & $417\pm1$ &   $495\pm2$  & $452\pm2$ \\ 
\hline 
 & (1) & $7044.24\pm 0.60$ & $6895.7\pm0.5$ & $6970.6\pm0.6$ \\ 
$K$ (m/s)     & (0) & $7032.27\pm0.68$ &   $6889.18\pm0.69$  & $6961.7\pm0.6$ \\
                & (0)+pl & $51.83\pm0.53$ &   $18.65\pm0.69$  & $26.2\pm0.7$ \\
\hline 
 & (1) & $0.23553\pm0.00007$  &  $0.24946\pm0.00008$  &  $0.25388\pm0.00008$  \\ 
$e$       & (0) & $0.23589\pm0.00009$ &   $0.24767\pm0.00011$  & $0.2524\pm0.0001$ \\
       & (0)+pl & $0.124\pm0.010$ &  $0.51\pm0.03$ &  $0.33\pm0.02$ \\
\hline 
$\dot{\omega}$ ($^\circ$/yr) & (1) & $-0.860\pm0.017$ & $-0.500\pm0.017$ & $-0.810\pm0.016$ \\ 
\hline \hline
 & (1) & 7.3  &  2.374  & 3.6 \\ 
$\sqrt{\chi^2}$    & (0) & 8.1  &  3.098  & 5.0 \\
                        & (0)+pl & 4.4  &  2.364 &  4.3       \\
\hline 
 & (1) & 36.3 & 13.54  & 19.3 \\ 
$rms$ (m/s)      & (0) & 39.1 & 16.92   & 26.7 \\
                    & (0)+pl & 22.8  &  13.30 &  22.8    \\
\hline \hline
\end{tabular}
\caption{Fits to $\nu$-Octantis (left) and simulations I and II (right). The radial velocity data in I and II was generated at the  221 observational data points for $\nu$-Octantis, covering  $t_{obs}=5.1$~y with precision about $5$~m/s.}
\end{minipage}
\end{table*}

\begin{figure}
  \centering
    \includegraphics[width=8cm]{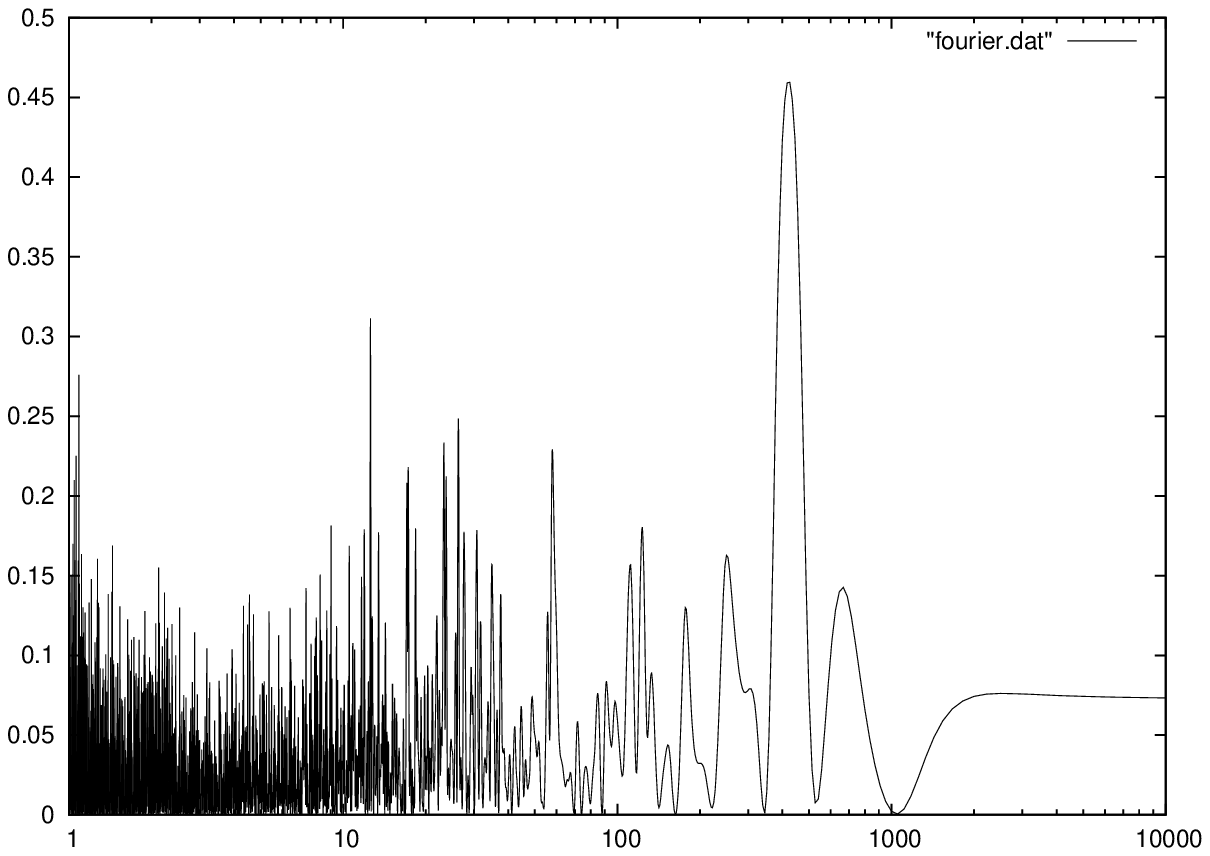}
    \includegraphics[width=8cm]{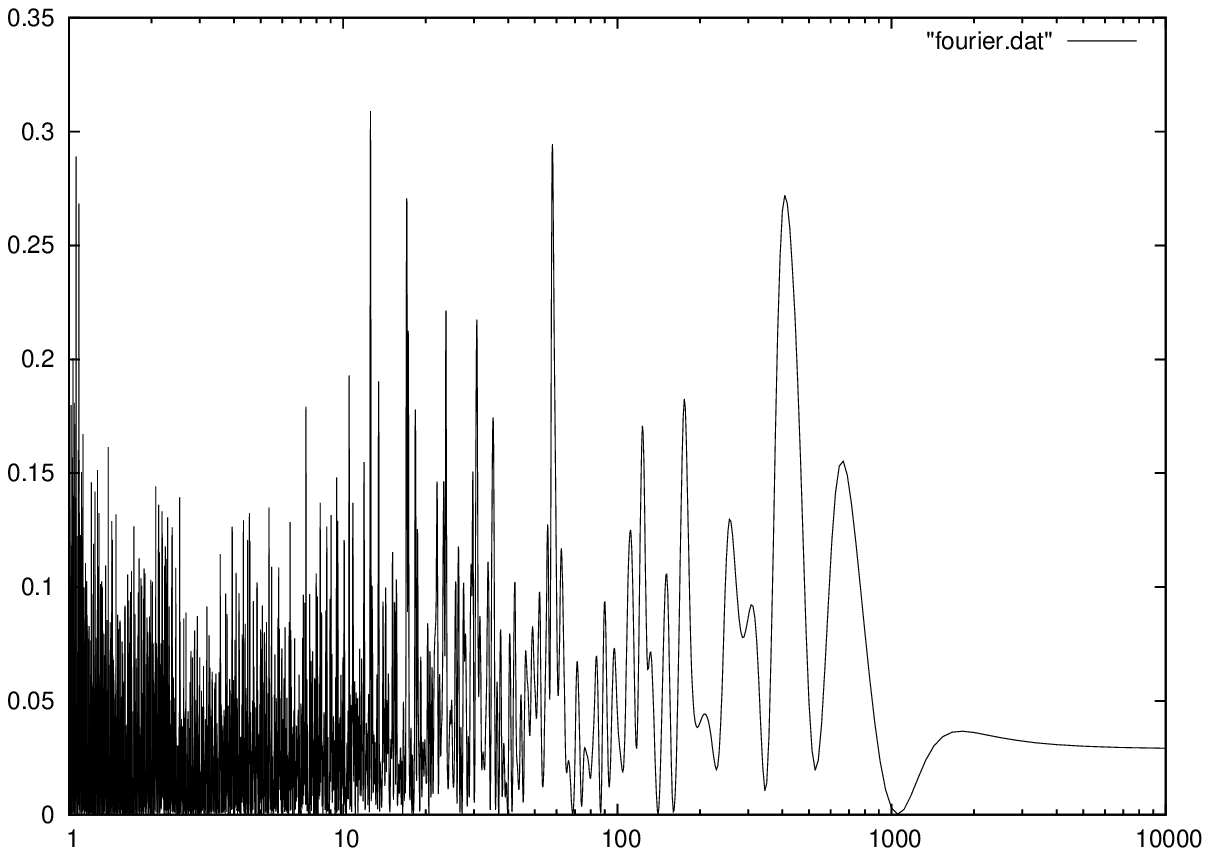}
  \caption{Fourier analysis of residuals after fit (0) and fit (1) to $\nu$-Octantis radial velocity data (top and bottom, respectively).}
  \end{figure}
 
We saw that retrograde precession occurs if the secondary star in $\nu$-Octantis is in turn a binary system inclined more than $45^\circ$ with respect to the main binary's orbit (Fig.~2).
In this scenario, we saw that a  periodogram of the residuals leftover from fitting a Keplerian orbit to the main binary could exhibit peaks that might be mistaken as planets. We saw that these peaks appeared close to harmonics of the main binary's orbital frequency. In particular, we showed an example (Sect. 4.3 and Fig.~4) where the
fake planet's  period is nearly commensurate (ratio 2/5) with the main binary's period which is exactly what happens in $\nu$-Octantis. Therefore, we propose that a hidden binary system could mimic a planet similar to the one reported in $\nu$-Octantis by \citet{Ramm_etal2009}. 

In order to estimate inner binary parameters that lead to retrograde precession of the outer binary at a rate $\dot{\omega}_2=-0.86^\circ/yr$, we set $a_1=a_c=0.35$~AU (Eq.~\ref{acritical}), $i=60^\circ$  and $e_1=0.76$  (i.e.\ the inner binary is at the Kozai stationary solution). Replacing these in Eqs.~(25) and (26) with $a_2=2.55$~AU, $m_2=1.4\,M_{\odot}$ and $m_b=0.5\,M_{\odot}$,  we obtain estimates for the inner binary's masses of $m_1=0.23\,M_{\odot}$ and $m_0=0.27\,M_{\odot}$. In Table~5 we show the results of fitting a  precessing Keplerian orbit (fit 1) and a fixed Keplerian orbit (fit 0) to such triple system (simulation I). We see that fit (1) is better than fit (0) and is comparable with the fit of a planet at 495~day.

We can also estimate inner binary parameters by performing N-body fits to $\nu$-Octantis radial velocity data. We assumed $m_2=1.4\,M_{\odot}$, and fixed the outer binary's orbit at $I_2=70.8^\circ$ and $\Omega_2=87^\circ$ which are the parameters inferred from the spectroscopic-astrometric solution (Table 4). We present the best fit solution in Table 6.  Retrograde precession occurs because $i=122^\circ$. Moreover, the best binary N-body fit (Table 6) is comparable to the Keplerian  planet fit (Table 4).  Although this solution is unstable because $a_1\approx 1.5\, a_c$, we can obtain "equivalent" stable configurations by reducing $a_1$, while maintaining $m_b=0.538\,M_{\odot}$ constant, and increasing the ratio $m_1/m_b$, so that the Keplerian term and the quadrupole interaction term are both kept constant.

\begin{table}
\centering
\begin{tabular}{|c|c|}
\hline \hline
\multicolumn{2}{c}{$\nu$-Octantis A}  \\ \hline 
\multicolumn{2}{c}{$m_A=1.4\,M_{\odot}$} \\ \hline \hline
\multicolumn{2}{c}{$\nu$-Octantis B = hidden binary system} \\ \hline      
$T_2=(1078\pm1)$~day & $T_1=(189.1\pm1.4)$~day \\
$K_2=(7010\pm3)$~m/s & $K_1=(2812\pm76)$~m/s \\
$e_2=0.2504\pm0.0003$ & $e_1=0.67\pm0.03$ \\
$\omega_2=(72.63\pm0.13)^\circ$ & $\omega_1=(23.25\pm1.38)^\circ$ \\
$I_2=(70.8\pm0.9)^\circ$ & $I_1=(63.4\pm2.7)^\circ$ \\
$\Omega_2=(87\pm1)^\circ$ & $\Omega_1=(232\pm1)^\circ$ \\  \hline
$m_0=0.496\,M_{\odot}$ &   $m_1=0.042\,M_{\odot}$ \\   
$a_2=2.565$~AU & $a_1=0.524$~AU \\     
\hline 
\multicolumn{2}{c}{$\sqrt{\chi^2}=4.9$}  \\
\multicolumn{2}{c}{$rms=25.9$~m/s}  \\
\hline \hline
\end{tabular}
\caption{Best N-body fit for $\nu$-Octantis assuming hierarchical  triple star system.}
\end{table}

We performed a simulation of such a stable configuration with $a_1=a_c=0.35$~AU, $m_1=0.109\,M_{\odot}$ and $m_0=0.429\,M_{\odot}$. In Table~5 we show the results of fitting a  precessing Keplerian orbit (fit 1) and a fixed Keplerian orbit (fit 0) to such triple system (simulation II). We see that fit (1) is better than fit (0) but fit (1) is better than the fit of a planet at 452~day.  The values of the precession rate after fit (1), and both $\sqrt{\chi^2}$ and residuals ($rms$) after fitting the planet's orbit are almost equal to those obtained for the real system $\nu$-Octantis (cf.~Table~4). 
 
From Table 5, as described above, we see that for simulated data the fit of a precessing Keplerian orbit is comparable (simulation I) or better (simulation II) than the fit of a planet, while in the real case ($\nu$-Octantis) the fit of a planet is currently better than the fit of a precessing Keplerian orbit. However, we stress that there are many more combinations of parameters ($m_0$, $m_1$, $i$ and $e_1$) that can cause a precession rate $\dot{\omega}_2=-0.86^\circ/yr$, assuming that it is well constrained.  Moreover, we expect our model to explain better the synthetic data generated with 3-body simulations than real data ($\nu$-Octantis) where we could have other planets or even stellar variability.
     
A planet around the primary star in $\nu$-Octantis  can also cause precession of the main binary's orbit.  However, our simulations show that  coplanar retrograde planet orbits, as reported in \citet{Eberle&Cuntz2010}, cause slow prograde precession of the main binary's orbit  at a rate $0.04^\circ$/yr. This is also what we expect from our quadrupole order theory (Eqs.~\ref{om2dotgen0},~\ref{atheta}) although we do not expect it to be accurate at semi-major  axis ratio $\alpha=0.47$. 
We saw (Sect.~2.3) that in order to have retrograde precession we would need the planet's orbit to be  inclined more than $45^\circ$ with respect to the $\nu$-Octantis binary. In our numerical integrations we could not find (although we did not do an exhaustive search) stable planet orbits at semi-major axis ratio $\alpha=0.47$ and with such high inclination with respect to the main binary. 

\section{Conclusion}

We studied the effect of a binary system on a nearby star's motion. This is a complement of our previous work \citep{Morais&Correia2008,Morais&Correia2011} where we assumed that we had observations for a fraction of the star's orbit around the binary's centre of mass.  Here, we assumed that we  had observations for a few orbits of the star around the binary's centre of mass.  We saw that, in this case, the secular effect of the binary dominates over the short-term effects.

We developed a secular theory which was based on a quadrupole expansion of the Hamiltonian. This is accurate for hierarchical triple systems composed of an inner binary, and a star that moves around this inner binary's centre of mass on a wider orbit which we called the outer binary.

We derived an expression for the outer binary's precession rate and showed that it is approximately constant. Therefore, the star's radial velocity can be modeled as a modified Keplerian radial velocity curve with slowly drifting amplitude.  We then showed how we can measure the outer binary's precession rate by fitting a precessing Keplerian orbit to the radial velocity data.  We also showed how we can estimate inner binary parameters from the measured precession rate.

We saw that, if we are unaware of the inner binary's existence and simply fit a non-precessing Keplerian orbit to the radial velocity data,  a  periodogram of the residuals will show peaks at or nearby harmonics of the outer binary's period which can be mistaken as planets.  However, if we fit a precessing Keplerian orbit to the radial velocity data, these signals  are no longer prominent in the leftover residuals. We conclude that detecting precession in the radial velocity data of a star within a binary system may be an indication that there is an unresolved third star.

We discussed the case of $\nu$-Octantis which is a close binary system ($2.55$~AU)  composed of a K-type star ($\nu$-Octantis A) and a fainter companion ($\nu$-Octantis B). Radial velocity data analysis showed a signal at $417$~day which was identified as a planet at $1.2$~AU of  $\nu$-Octantis A. However, we showed that the radial velocity data  currently implied  retrograde precession of about $-0.86^\circ$/yr for this binary system. We suggested that this may indicate that $\nu$-Octantis B is actually a double star which could explain a signal similar to that previously associated with a planet.
At the moment we cannot yet decide that the reported planet of $\nu$-Octantis A is simply an artifact caused by $\nu$-Octantis B being a double star. 
In order to distinguish between the two hypothesis (planet or double star), more radial velocity data for $\nu$-Octantis is needed, so that we can better constrain the main binary's precession rate.  Moreover,  the planet hypothesis could be compatible with retrograde precession of the $\nu$-Octantis binary if we could  prove the existence of stable orbits around $\nu$-Octantis A, with semi-major axis ratio  $\alpha=0.47$ and inclined more than $45^\circ$ with respect to the $\nu$-Octantis binary\footnote{As we were in the process of  submitting the final version of this article, we became aware of a stability study done by K. Gozdziewski which confirms that stable planetary solutions compatible with $\nu$-Octantis radial velocity data are unlikely.}.

\subsection*{Acknowledgements}
We acknowledge financial support from FCT-Portugal (grant PTDC/CTE-AST/098528/2008).

\bibliographystyle{mn2e}

\bibliography{closebinary}

\end{document}